\documentclass[twocolumn,prl,nofootinbib,preprintnumbers,superscriptaddress]{revtex4}
\usepackage{multirow}
\usepackage{amsmath}
\usepackage{epsfig}
\usepackage{subfigure}
\usepackage{float}
\usepackage{verbatim} 
\usepackage{booktabs}
\usepackage[Q=yes,pverb-linebreak=no]{examplep}

\newcommand{\mgamcatnlo}{{\sc MadGraph5\Q{_}aMC@NLO}}
\newcommand{\madgraph}{{\sc MadGraph}}
\newcommand{\madloop}{{\sc MadLoop}}
\newcommand{\cuttools}{{\sc CutTools}}
\newcommand{\madfks}{{\sc MadFKS}}
\newcommand{\openloop}{{\sc OpenLoop}}

\begin{document}

\preprint{CERN-TH/2013-276}

\title{The top quark induced backgrounds to Higgs production in the $WW^{(*)}
  \to ll\nu\nu$ decay channel at NLO in QCD}

\author{R. Frederix}
\affiliation{PH Department, TH Unit, CERN, CH-1211 Geneva 23, Switzerland}

\begin{abstract}
We present the complete NLO contributions to the $pp \to e^+ \nu_e
\mu^- \bar{\nu}_{\mu} b \bar{b}+X$ process in the four flavour scheme,
i.e.~with massive b quarks, and its contribution to the $H \to
WW^{(*)} \to ll\nu\nu$ measurement in the 1-jet bin at the LHC. This
background process includes top pair, single top and non-top
quark-resonant contributions. The uncertainty at NLO from
renormalisation and factorisation scale dependence is about
${+30}\%\,\,{-20}\%$. We show that the NLO corrections are relatively
small, and that separating this background in top pair, $Wt$ and
$b$-quark associated $ll\nu\nu$ production is a fair approximation.
\end{abstract}

\maketitle
After the Higgs boson discovery~\cite{Aad:2012tfa,Chatrchyan:2012ufa}
the next step is to determine whether this boson is (solely)
responsible for the electroweak symmetry breaking and for generating
the masses of all the fundamental particles. The way to tackle this is
to extract the coupling constants of the Higgs boson by measuring the
Higgs boson cross sections in as many production and
decay modes as possible.

For the discovery of the Higgs boson the three most important channels
were the $H \to \gamma\gamma$, $H \to ZZ^{(*)} \to 4l$ and $H\to
WW^{(*)} \to ll\nu\nu$ decay modes. Even though the latter has the
largest branching ratio, it has the smallest contribution to the Higgs
signal significance. This comes as no surprise: due to the presence of
two neutrinos in the final state, the reconstruction of the Higgs
signal in the form of a narrow resonance peak over a flat background
is not possible for this decay mode. This makes the separation of the
Higgs signal from (non) reducible backgrounds much more complicated
and precise predictions for the backgrounds are needed to determine
the excess of events that can be attributed to the Higgs signal.

To increase the significance in the extraction of the Higgs
contribution for the $H\to WW^{(*)} \to ll\nu\nu$ channel, the data is
separated in jet bins by the CMS and ATLAS
experiments~\cite{Chatrchyan:2012ty,ATLAS-CONF-2013-030}.  In the
0-jet bin, the dominant background is the non-reducible $pp\to WW$
production.  In the 1-jet bin, where each event is required to have
exactly 1 jet in association with the two charged leptons and the
missing $E_T$, also the backgrounds from top quarks are large; mostly
top pair and $Wt$ production. For a reliable simulation of these
backgrounds, including next-to-leading order (NLO) QCD corrections in
the calculation is essential.  In this letter, we present the top
induced background to Higgs production in the 1-jet bin, without
separating top pair and $Wt$ production and thus keeping all their
interference effects. This requires the calculation of the NLO
corrections to the $pp \to e^+ \nu_e \mu^- \bar{\nu}_{\mu} b
\bar{b}+X$ process in the four-flavour (4F) scheme\footnote{Results
  for this process have also recently been presented by S.~Kallweit at
  the RADCOR symposium, see
  \href{www.ippp.dur.ac.uk/RADCOR2013/}{www.ippp.dur.ac.uk/RADCOR2013/}.},
keeping the $b$ quark mass finite, which we present here for the first
time.

The NLO corrections to the $pp \to e^+ \nu_e \mu^- \bar{\nu}_{\mu} b
\bar{b}+X$ process in the \emph{five-flavour (5F) scheme} are
known~\cite{Denner:2010jp,Bevilacqua:2010qb,Denner:2012yc}. In
the 5F scheme the mass of the $b$ quark is neglected, which
means that the above process is not finite in fixed-order perturbation
theory without requiring phase-space cuts on the final state $b$
jets. Therefore, such a calculation cannot be used to estimate the
$Wt$ and top pair production processes and, moreover, it cannot be
used to estimate the top background in the 1-jet bin in the $H\to
WW^{(*)} \to ll\nu\nu$ measurement, where a veto on a second jet is
needed.

\begin{figure}[h]
\centering
\subfigure[]{\includegraphics[scale=.30]{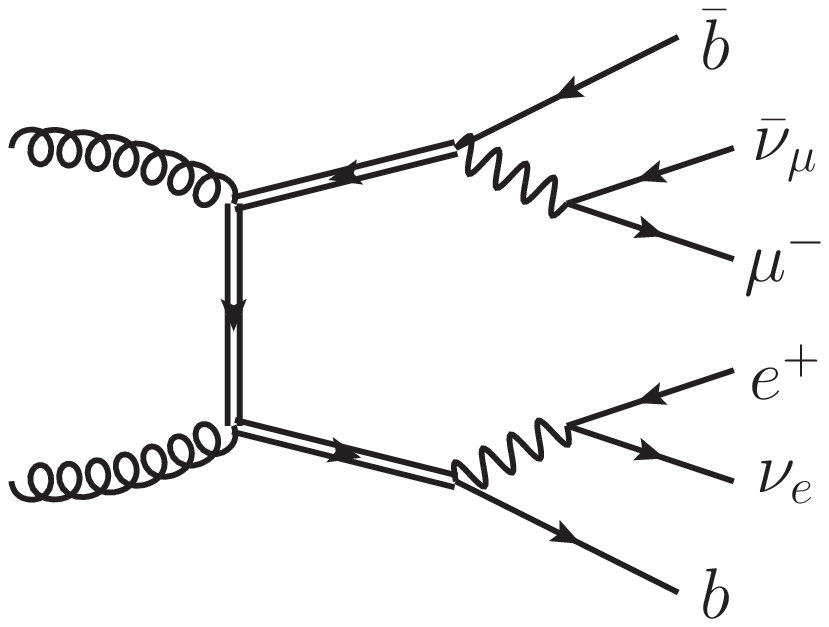}}
\subfigure[]{\includegraphics[scale=.30]{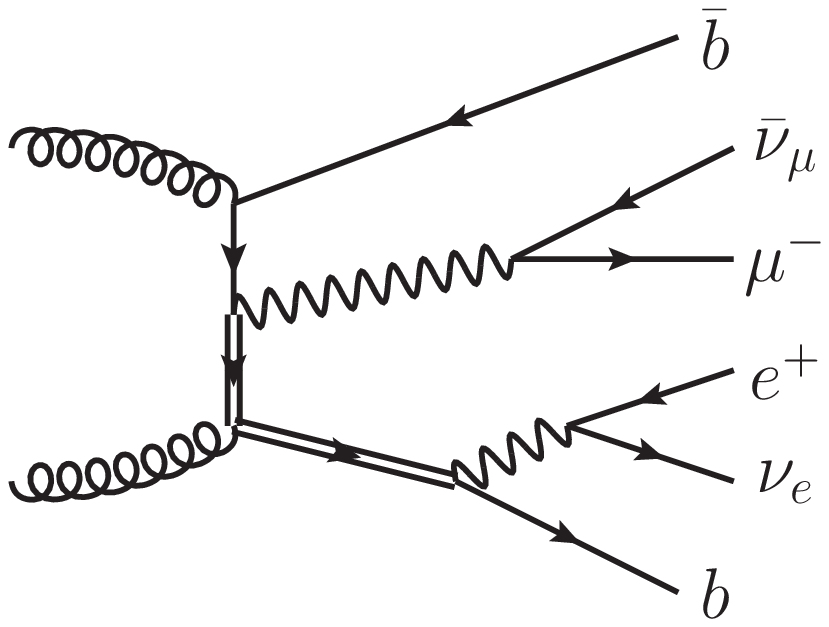}}
\subfigure[]{\includegraphics[scale=.30]{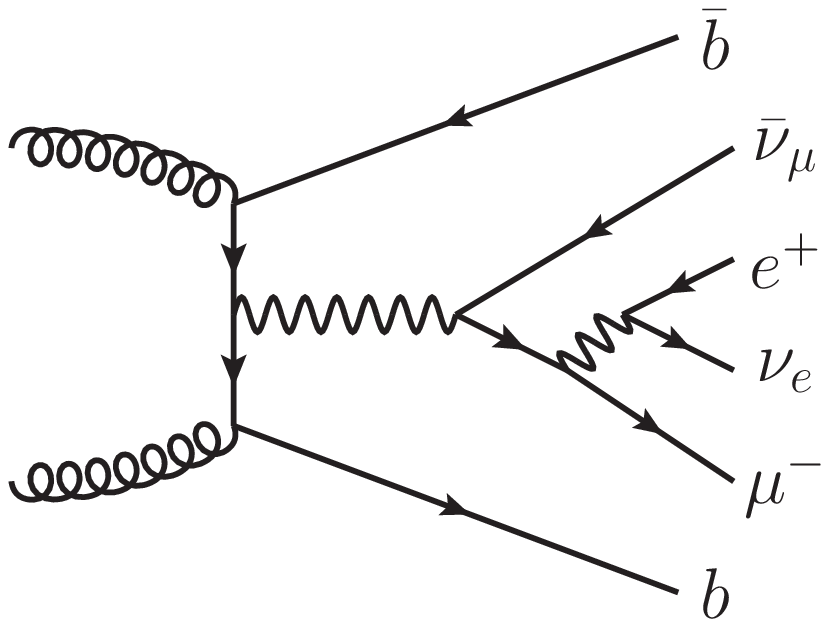}}
\caption{Representative LO diagrams for top pair (a), $Wt$ (b) and
  $b$-quark associated $ll\nu\nu$ (c) contributions to $pp \to e^+ \nu_e
  \mu^- \bar{\nu}_{\mu} b \bar{b}+X$ production. Top quarks are denoted by
  double fermion lines.}
\label{diagrams}
\end{figure}

The calculation of the $pp \to e^+ \nu_e \mu^- \bar{\nu}_{\mu} b
\bar{b}+X$ process in the 4F scheme includes double
top-quark resonant production (``top pair production''), single
top-quark resonant contributions (``$W$ boson associated single top
production'') as well as non top-quark resonant contributions (
``$b$-quark associated $ll\nu\nu$ production''). In
Fig.~\ref{diagrams} three representative LO Feynman diagrams are shown
for this process. The calculation includes all the interference
effects between the various contributions, as well as all off-shell
effects. In the 4F scheme the $b$ quarks are treated as
massive particles, the running of the strong coupling is performed
with four flavours and a 4F PDF set should be used. Keeping
the $b$ quark massive in the calculation implies that even in the
absence of any phase-space cuts, the perturbative expansion yields
finite results. For the NLO computation presented here, the complete
$\mathcal{O}(\alpha_s)$ corrections have been included without
resorting to any approximations.

The calculation has been performed within the
\mgamcatnlo~framework~\cite{mg5_amcatnlo}: the diagram generation is
done by \madgraph~\cite{Alwall:2011uj}, the one-loop corrections are
obtained with \madloop~\cite{Hirschi:2011pa}, which is based on the
OPP reduction method~\cite{Ossola:2006us} and its implementation in
\cuttools~\cite{Ossola:2007ax} and uses the
\openloop~algorithm~\cite{Cascioli:2011va}. The phase-space
integration and the cancellation of IR divergences in intermediate
steps of the calculation are dealt with by
\madfks~\cite{Frederix:2009yq}, which is based on the FKS subtraction
method~\cite{Frixione:1995ms}. Furthermore, we use the complex mass
scheme~\cite{Denner:1999gp,Denner:2005fg}, as implemented in
\mgamcatnlo~\cite{ComplexMassScheme}, for the treatment of the widths
of heavy resonances.
The implementation of the
various contributions in the \mgamcatnlo~program is done in a
completely process-independent way and allows for complete automation
of the calculation with extensively tested algorithms, which
minimises the possibility of computational errors.

\begin{figure}[ht!]
\centering
\includegraphics[scale=.55]{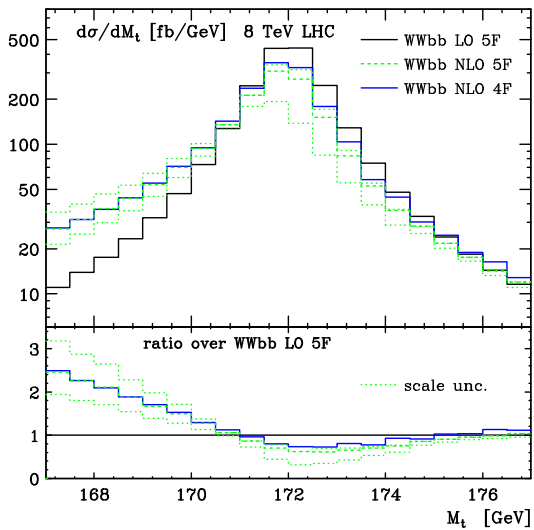}
\caption{Top quark invariant mass computed by taking the invariant
  mass of the positron, electron-neutrino and the jet containing the
  $b$ quark, using the same parameters and cuts as in
  Ref.~\cite{Denner:2012yc} for the $pp \to e^+ \nu_e \mu^-
  \bar{\nu}_{\mu} b \bar{b}+X$ process. The LO (black solid) and NLO
  (green dashed) predictions in the 5F scheme and NLO in the 4F scheme
  (blue solid) are plotted. The green dotted lines show the scale
  uncertainty on the NLO 5F result.}
\label{mt}
\end{figure}

As a further check on the implementation of the various algorithms in
the \mgamcatnlo~program, we have compared the results obtained in the
5F scheme by Denner~\textit{et al.}~in Ref.~\cite{Denner:2012yc} by a
5F scheme calculation obtained within our framework and found
excellent agreement. As an example, in Fig.~\ref{mt} we show the
positron--electron-neutrino--$b$-jet invariant mass, using exactly the
same cuts and parameters as used in Ref.~\cite{Denner:2012yc}; thus
this plot can be directly compared to Fig.~16~of that reference. In
the plot, we also present the same observable simulated in the 4F scheme,
i.e.~the calculation presented for the first time in this letter. The
4F and 5F scheme results lie (almost) exactly on top of each other for
$M_t<172\,\textrm{GeV}$, where the shape is dominated by resonant
contributions ($M_t\simeq 172\,\textrm{GeV}$) in which some radiation
from the bottom quark is not clustered in the $b$ jet. The region $M_t
> 172\,\textrm{GeV}$ is more sensitive to the exact treatment of the
(non)-resonant contributions~\cite{Papanastasiou:2013dta}, but also
here the differences between 4F and 5F scheme calculation are small
and their respective scale uncertainty bands overlap. (Uncertainty
band for the 4F scheme calculation is not shown).

To predict the top induced backgrounds in the $H \to WW^{(*)}\to
ll\nu\nu$ decay channel at the 8 TeV LHC we use the following
parameters. The mass of the top quark is set to
$m_t=172\,\textrm{GeV}$, the mass of the bottom quark is
$m_b=4.7\,\textrm{GeV}$ and the mass of the $W$ and $Z$ bosons are
$m_W=80.399\,\textrm{GeV}$ and $m_Z=91.188\,\textrm{GeV}$. The
(inverse of the) weak coupling is set to $\alpha^{-1}=132.35$. The
width of the $W$ and $Z$ bosons are $\Gamma_W=2.100\,\textrm{GeV}$ and
$\Gamma_Z=2.510\,\textrm{GeV}$, respectively. At LO the top width is
set to $\Gamma_t^{\textrm{LO}}=1.44\,\textrm{GeV}$ and at NLO we use
$\Gamma_t^{\textrm{NLO}}=1.32\,\textrm{GeV}$. For the description of
the partons inside the colliding protons, the 4F
\texttt{MSTW2008nf4(N)LO} PDF sets~\cite{Martin:2009iq} are used for
the (N)LO calculation. This also defines the numerical value and
running of the strong coupling constant.

We impose the following ``Higgs measurement'' cuts on the final state
particles, which are are motivated by the cut-based analysis for the
ATLAS measurement of the Higgs boson in the $WW^{(*)}\to ll\nu\nu$
decay channel~\cite{ATLAS-CONF-2013-030}. We require two charged
leptons in the central rapidity region, $|\eta_l|<2.4$, of which the
hardest needs to have a transverse momentum $p_T^{\textrm{lead}} > 25 $
GeV and the other a $p_T^{\textrm{sublead}} > 15 $ GeV. The lepton
invariant mass should be larger than $m_{ll} > 10 $ GeV. We require
exactly one jet with $|\eta_j|<4.7$, defined using the anti-$k_T$
algorithm~\cite{Cacciari:2008gp} with $\Delta R=0.5$ and a minimal
transverse energy of $E_T^j>30$ GeV, where
$E_T=\sqrt{m^2+p_T^2}$.
Furthermore, there should be a sizeable (relative) missing transverse
energy $E_{T,\textrm{rel}}^\textrm{miss} > 25$ GeV, where
$E_{T,\textrm{rel}}^\textrm{miss}=s\,E_T^\textrm{miss}$, with
$E_T^\textrm{miss}$ the usual missing transverse energy and
$s=\sin|\Delta\phi_{\textrm{closest}}|$, i.e.~the sinus of the
azimuthal separation between the missing transverse momentum vector
and the closest charged lepton or the jet if there is one in the same
azimuthal hemisphere. If there is no such object, $s=1$. The MC truth
is used to define the missing transverse energy vector as the
(transverse part of the) sum of the two neutrino momenta. Finally, the
following $H\to WW^{(*)}\to l\nu l\nu$ topology cuts are applied: the
lepton invariant mass should be smaller than $m_{ll} <50 $ GeV and
their azimuthal separation smaller than $|\Delta\phi_{ll}|<1.8$.

Due to the complexity of the process, and the rather involved set of
cuts, it is not straight-forward to define a single hard scale for
this process that could be used as a renormalisation and factorisation
scale. We have therefore chosen a scale that is not very specific to
this process, but should capture well the general hardness of the
kinematics. This central scale is $\mu_R^0=\mu_F^0=H_T/2$,~i.e.~half
the scalar sum of the transverse energies of all the final state
particles/partons (including the two neutrinos). With this central scale,
NLO corrections are relatively small for both the inclusive process as
well as after applying the cuts described above. To assess
contributions from beyond NLO we assign an uncertainty to our predictions
by computing the envelope of the results with renormalisation and
factorisation scales equal to
$(\mu_R,\mu_F)=\{(1,1),(0.5,0.5),(2,2),(0.5,1),(2,1),(1,0.5),(1,2)\}\times
(\mu_R^0,\mu_F^0)$. These 7 values are obtained at no extra CPU cost
using the reweighting method described in Ref.~\cite{Frederix:2011ss}.

\begin{figure}[t!]
\centering
\includegraphics[scale=.55]{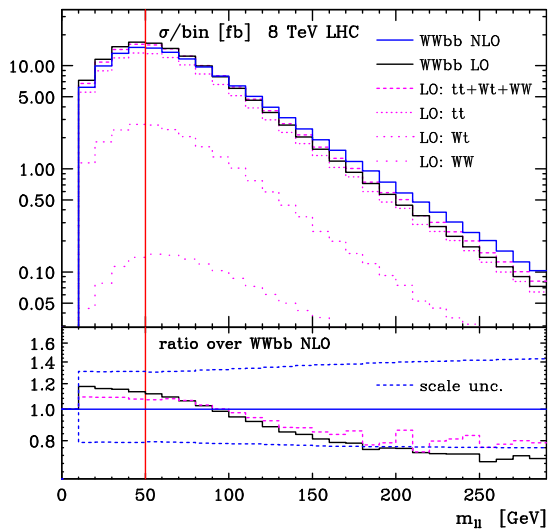}
\caption{Invariant mass of the charged lepton pair for the $pp \to e^+
  \nu_e \mu^- \bar{\nu}_{\mu} b \bar{b}+X$ process in the 4F
  scheme, with the Higgs measurement cuts, apart from the cut on the
  charged lepton invariant mass $m_{ll}<50$ GeV.}
\label{mll}
\end{figure}

\begin{figure}[t!]
\centering
\includegraphics[scale=.55]{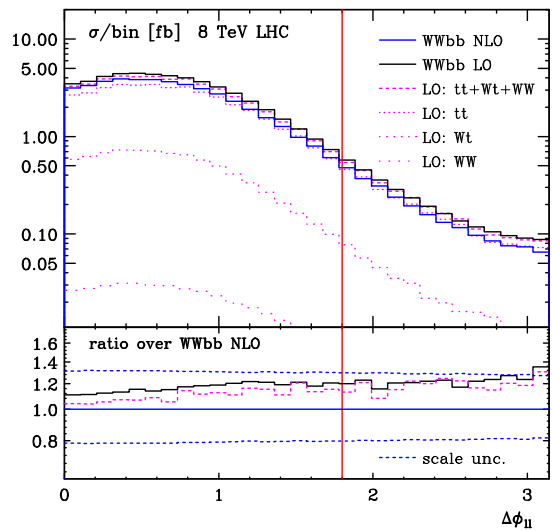}
\caption{Azimuthal separation of the charged leptons for the $pp \to
  e^+ \nu_e \mu^- \bar{\nu}_{\mu} b \bar{b}+X$ process in the 4F
  scheme, with the Higgs measurement cuts, apart from the cut on the
  charged lepton invariant mass $|\Delta\phi_{ll}|<1.8$.}
\label{dphill}
\end{figure}

\begin{figure}[t!]
\centering
\includegraphics[scale=.55]{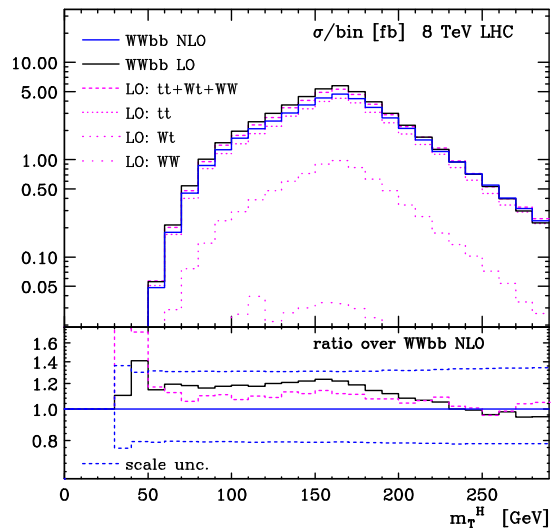}
\caption{Higgs transverse mass for the $pp \to e^+ \nu_e \mu^-
  \bar{\nu}_{\mu} b \bar{b}+X$ process in the 4F scheme, with
  the Higgs measurement cuts.}
\label{mth}
\end{figure}

In Figs.~\ref{mll}-\ref{mth} we show the invariant mass of the two
charged leptons ($m_{ll}$), the azimuthal separation of the two
leptons ($\Delta\phi_{ll}$) and the transverse mass of the Higgs boson
($m_T^H$), respectively. The latter is defined as
$m_T^H=\sqrt{(E_T^{ll}+E_T^{\textrm{miss}})^2 -
  |\mathbf{p}_T^{ll}+\mathbf{E}_T^{\textrm{miss}}|^2 }$, where
$E_T^{ll}=\sqrt{|\mathbf{p}_T^{ll}|^2+m_{ll}^2}$. The $m_{ll}$ and
$\Delta\phi_{ll}$ variables are used to define the ``Higgs topology''
cuts, while the $m_T^H$ distribution is used to extract the Higgs
signal in the cut-based analysis by ATLAS~\cite{ATLAS-CONF-2013-030}.
In the plots, results for the full $pp \to e^+ \nu_e \mu^- \bar{\nu}_{\mu}
b \bar{b}+X$ process at LO (labelled ``\texttt{WWbb LO}'') and NLO
(``\texttt{WWbb NLO}'') are presented. Also shown are the separate LO
calculations for top pair production (``\texttt{LO: tt}''), $W$-boson
associated single top production (``\texttt{LO: Wt}''), $b$-quark
associated $ll\nu\nu$ production (``\texttt{LO: WW}'') and their sum
(``\texttt{LO: tt+Wt+WW}''). These latter processes are defined in the
narrow width approximation,~i.e.~in the \texttt{LO: tt} process we
take only diagrams with two s-channel top quark propagators into
account (e.g.~Fig.~\ref{diagrams}(a)), \texttt{LO: Wt} has only
diagrams with one s-channel top quark propagator
(e.g.~Fig.~\ref{diagrams}(b)), while the \texttt{LO: WW} process has
no s-channel top quark propagators in any of its contributing diagrams
(e.g.~Fig.~\ref{diagrams}(c)); all other parameters are the same as
used for the \texttt{WWbb LO} predictions. The differences between the
\texttt{LO: tt+Wt+WW} and \texttt{WWbb LO} results stem only from
interference effects (among the three contributions to the \texttt{LO:
  tt+Wt+WW} prediction) which are only included in the complete
\texttt{WWbb} simulations.
For the $m_{ll}$ and $\Delta\phi_{ll}$ plots, the Higgs topology cut
on that distribution is not applied, and is denoted by the vertical
line. 
In the lower inset of the plots, the ratio is
taken~w.r.t.~the \texttt{WWbb NLO} result. Also shown here is the
relative scale uncertainty on the NLO result. For the observables
studied here, LO scale uncertainties are marginally larger than the
NLO scale uncertainties. We refrain from showing these, because scale
uncertainties at LO are not a proper estimate of missing higher order
corrections, in particular when jet veto's are applied or when studing
exclusive jet bins, as is done here.

As can be seen from the plots, the NLO corrections are small and
negative (about -20\%) when all the cuts are applied. This is within
the NLO scale uncertainty, which is about ${+30}\%\,\,{-20}\%$. It is
no surprise that the scale dependence at NLO is still sizeable because
the observables considered are rather exclusive; particularly due to
the veto of any jets beyond the hardest. There is a visible difference
in shape between the LO and NLO results, even though it stays within
the NLO scale uncertainties. This difference in shape in the $m_{ll}$
distribution is of particular importance, because the region at large
invariant mass, $m_{ll}>80\,\textrm{GeV}$, is used as a background
control sample in the ATLAS analysis.

When summing the separate LO calculations for $t\bar{t}$, $Wt$ and
$b$-quark included $WW$ production, the final result is very close to
the \texttt{WWbb LO} prediction. The bulk is given by top pair
production, $Wt$ being about a factor 5 smaller and the contribution
from $b$-quark associated $WW$ production is approximately 2 orders of
magnitude smaller than top pair production. Interesting to see is that
the sum of the separate calculations lies in between the LO and NLO
\texttt{WWbb} results. This is not only the case for the 3 observables
presented here, but seems to be a general feature of this
process. Most likely, this is due to positive interference effects
between $t\bar{t}$ and $Wt$ production, which results in a slightly
higher cross section when these effects are taken into account. On top
of that, the NLO corrections bring the results down again, slightly
below the LO result without interference effects.

To conclude, we have presented a calculation of the top quark induced
backgrounds to the $H \to WW^{(*)} \to ll\nu\nu$ measurement in the
1-jet bin by computing the NLO corrections to the process $pp \to e^+
\nu_e \mu^- \bar{\nu}_{\mu} b \bar{b}+X$ in the four-flavour
scheme. This process yields a consistent description of top pair
production, $W$-boson associated single top production and $b$-quark
associated $WW$ production, including all interference and off-shell
effects. Using $H_T/2$ as a central renormalisation and factorisation
scale, the NLO corrections are small and negative in the Higgs signal
region and its uncertainties from renormalisation and factorisation
scale dependence are about ${+30}\%\,\,{-20}\%$. There is, however, a
difference in shape between LO and NLO in the $m_{ll}$ distribution:
the negative NLO corrections at small $m_{ll}$ (Higgs signal region)
turn positive for large $m_{ll}$, which is the background control
region. Therefore, this might have an effect on the recent Higgs boson
measurements and on future high-precision extractions of its coupling
constants.  Using separate calculations for $t\bar{t}$ and $Wt$ (and
$b$-quark associated $WW$) production based on the narrow width
approximation for the top quark, yields a fair approximation of the
final results, within left-over theoretical uncertainties.

For a more exclusive description of the final state, matching to the
parton shower would be required. This would allow for an improved
description of the jets, in particular when b-tagging would be
applied, and a fair comparison to the data can be made. When matching
to the parton shower using the MC@NLO
technique~\cite{Frixione:2002ik}, care has to be taken to prevent
double counting from the NLO corrections and parton shower
(non-)emissions from the intermediate top quarks. Work in this
direction has already been started.

I would like to thank V.~Hirschi for writing the
\texttt{virt\Q{_}reweighter.py} code that has been used to obtain and
check some parts of the results presented here. I also thank
S.~Frixione, F.~Maltoni, V.~Hirschi, O.~Mattelaer, M.~Zaro and
P.~Torrielli for proof-reading this manuscript. 

\bibliography{wwbb}

\end{document}